\documentclass[
 reprint,
 showpacs,
 amsmath,amssymb,
 aps
]{revtex4-1}
\usepackage{subfigure}
\usepackage{graphicx}
\usepackage{dcolumn}
\usepackage{bm}
\usepackage{color}
\usepackage[colorlinks,linkcolor=red,anchorcolor=blue,citecolor=green]{hyperref}
\usepackage{rotfloat}
\usepackage{cleveref}
\usepackage{natbib}
\def \bea{\begin{eqnarray}}
\def \eea{\end{eqnarray}}

\def \ba{\begin{array}{cccc}}
\def \ea{\end{array}}

\begin{document}
\title{A programmable $\bm{k\cdot p}$ Hamiltonian method and application to magnetic topological insulator MnBi$_2$Te$_4$}
\author{Guohui Zhan$^{1}$, Minji Shi$^{1}$, Zhilong Yang$^{1}$, Haijun Zhang$^{1,2\ast}$}
\affiliation{
 $^1$ National Laboratory of Solid State Microstructures, School of Physics, Nanjing University, Nanjing 210093, China\\
 $^2$ Collaborative Innovation Center of Advanced Microstructures, Nanjing University, Nanjing 210093, China
 }
\email{zhanghj@nju.edu.cn}
\date{\today}

\begin{abstract}
In the band theory, first-principles calculations, the tight-binding method and the effective $\bm{k\cdot p}$ model are usually employed to investigate the electronic structure of condensed matters.  The effective $\bm{k\cdot p}$ model has a compact form with a clear physical picture, and first-principles calculations can give more accurate results. Nowadays, it has been widely recognized to combine the $\bm{k\cdot p}$ model and first-principles calculations to explore topological materials. However, the traditional method to derive the $\bm{k\cdot p}$ Hamiltonian is complicated and time-consuming by hand. In this work, we independently developed a programmable algorithm to construct effective $\bm{k\cdot p}$ Hamiltonians for condensed matters. Symmetries and orbitals are used as the input information to produce the one-/two-/three-dimensional $\bm{k\cdot p}$ Hamiltonian in our method, and the open-source code can be directly downloaded online. At last, we also demonstrated the application to MnBi$_2$Te$_4$-family magnetic topological materials.
\end{abstract}


\maketitle

Recently, the study of topological states and topological materials, such as, topological insulators, topological semimetals and topological superconductors, has become an important topic in condensed matter physics, and  great progresses have been achieved in both experiments and theories\cite{hasan2010,qi2011,armitage2018}. The developed topological band theory with first-principles calculations played a key role in the past years. For example, topological invariants, such as $\mathcal{Z}_2$, were defined through the band theory\cite{fu2007}. The concept of band inversion of topological states was birth from the band theory\cite{bernevig2006, zhang2012a}. Also, the topological boundary states were successfully predicted through first-principles calculations\cite{zhang2009}. However, the first-principles band structures seem like a black box with much-hidden information, which results in a barrier to deeply understand the essential physical pictures of topological states. Differently, the $\bm{k\cdot p}$ Hamiltonian has a simple form and a clear physical picture, which is a necessary and useful supplement for the first-principles calculations\cite{luttinger1955,kane1966,voon2009,liu2010}. With the $\bm{k\cdot p}$  Hamiltonian, Fu predicted an unconventional hexagonal warping term in surface states of topological insulator Bi$_2$Te$_3$\cite{fu2009}. Therefore, the combination of first-principles calculations and the $\bm{k\cdot p}$ Hamiltonian has become a standard paradigm to theoretically study  topological states and topological materials \cite{zhang2012a,zhang2009,xu2011}.

\begin{figure}[ht]
\includegraphics[width=9.2cm]{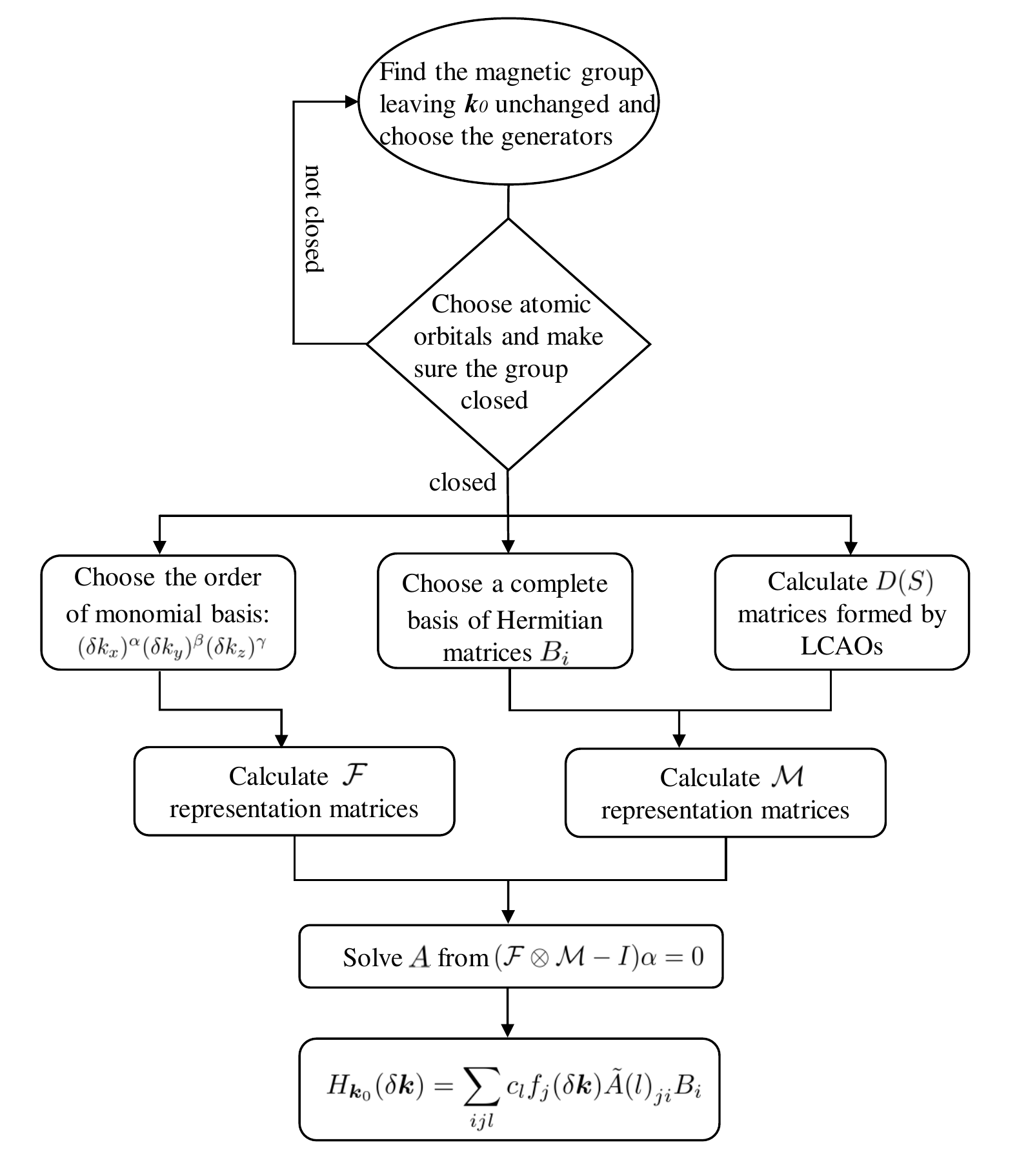}
\caption{\label{fig:fig1} \textbf{Workflow for constructing the $\bm k \cdot \bm p$  effective Hamiltonian}. The selected generators of the group and atomic orbitals are the input information to run the program. In the process, the key point is getting the representation matrix $\mathcal{F}$ and $\mathcal{M}$. By solving $(\mathcal{F}\otimes\mathcal{M}-I)\alpha=0$, the Hermitian matrices can be obtained from the solution $A$. }
\end{figure}

 Based on the group theory and the $\bm{k\cdot p}$ perturbation theory, Liu {\it et al}. provided an instructive demonstration showing how to construct the $\bm{k\cdot p}$ Hamiltonian of Bi$_2$Se$_3$-family three-dimensional topological insulators\cite{liu2010}. However, though the derivation is a standard method, the detailed process is quite troublesome and time-consuming by hand, especially for dealing with high-order terms of $\bm{k\cdot p}$ Hamiltonians. Therefore, it is practicable and highly necessary to develop automated programmable methods to efficiently construct $\bm{k\cdot p}$ Hamiltonians. In this context, there have been some innovative proposes, such as, {\it kdotp-symmetry} code developed by Gresch $et$ $al.$\cite{gresch2018,kdotp}, {\it Qsymm} Python package developed by Varjas $et$ $al.$\cite{varjas2018, qsymm}. In this letter, based on group theory, we independently developed a programmable algorithm to construct $\bm{k}\cdot\bm{p}$ Hamiltonians for all one-/two-/three-dimensional materials (the open-source code can be download from \url{https://github.com/shimj/Model-Hamiltonian}). With this automated algorithm, only the crystal symmetry and atomic orbitals involved are needed to produce the $\bm{k\cdot p}$ Hamiltonian, which effectively avoids time-consuming calculations and latent mistakes. The schematic of our programmable method is shown in Fig. \ref{fig:fig1}. We have successfully applied our method to MnBi$_2$Te$_4$-family magnetic topological  materials\cite{zhang2019,zhang2020,wang2020}.


\textit{Basic $\bm k\cdot\bm p$ Hamiltonian theory.} In the band theory, the wavefunction in a periodic lattice is written as the Bloch wavefunction, $\psi_{\bm{k}}(\bm{r})=u_{\bm{k}}(\bm{r})e^{i\bm{k}\cdot\bm{r}}$. Here, $u_{\bm{k}}(\bm{r})$ is a periodic function with $u_{\bm{k}}(\bm{r})=u_{\bm{k}}(\bm{r}+\bm{R})$ where $\bm{R}$ indicates the lattice vector. By substituting the Bloch wavefunction $\psi_{\bm{k}}$ into the Schr\"{o}dinger equation, we have
\begin{equation}
\mathcal{H}_{\bm k}u_{\bm k} = E_{\bm k}u_{\bm k},
\end{equation}
where the Hamiltonian has the $\bm{k}$-dependent form $\mathcal{H}_{\bm k}=p^2/2m+V+\hbar\bm{k\cdot p}/m_0+\hbar^2k^2/2m$. Generally, the representation matrix of Hamiltonian $\mathcal{H}_{\bm k}$ near a $\bm{k_0}$, with $\bm{k}=\bm{k_0}+\delta\bm{k}$, can be approached by Taylor series as
\begin{equation}\label{hnk}
\mathcal{H}_{\bm{k_0}+\delta\bm{k}}\doteq\sum_{\alpha+\beta+\gamma\leq n} (\delta k_x)^{\alpha}(\delta k_y)^{\beta}(\delta k_z)^{\gamma}\Gamma_{\alpha\beta\gamma},
\end{equation}
where $\Gamma_{\alpha\beta\gamma}$ are constant matrices, $n$ indicates the order of Taylor series expansion. In the following, we will derive the representation matrix of $\mathcal{H}_{\bm k}$ from the commutation relation between Hamiltonian and symmetry operations. 

When a periodic condensed matter system preserves a $S$ symmetry, its Hamiltonian $\mathcal{H}$ satisfies the commutation relation $[\mathcal{H},S]=0$. In $\bm{k}$-space, we have
\begin{equation}\label{Hk_final}
S\mathcal{H}_{\bm k}S^{-1} = \mathcal{H}_{\gamma\bm k},
\end{equation}
where \begin{equation}\label{equ:total_con}
\gamma = \begin{cases}
\tilde{g} & S = Q\\
-1 & S = \mathcal{T}\\
-\tilde{g} & S = Q\mathcal{T}
\end{cases}.
\end{equation}
Here, $Q$ represents an operator corresponding to the spatial operation $g=\{\tilde{g} | \tau\}$ and $\mathcal{T}$ (= $(-i\sigma_y\otimes I) \mathcal{K}$) is the time reversal operator,
$\tilde{g}$ is a rotation or reflection operator and $\tau$ is a translation vector. It is noteworthy that $2\pi$-rotation for a spinful system is not included in our discussion, since it leads to an identity $(-1)\mathcal{H}_{\bm k}(-1)=\mathcal{H}_{\bm k}$ without giving any constraint to the Hamiltonian $\mathcal{H}_{\bm k}$. In principle, any complete set of lattice-periodic functions, for example $\{e^{i\bm G\cdot \bm r}\}$, can be used as basis to accurately solve the Schr\"{o}dinger equation $\mathcal{H}_{\bm k}|u_{\bm k}\rangle = E_{\bm k}|u_{\bm k}\rangle$. Since we are interested in the bands around a certain point $\bm k_0$ in the Brillouin Zone (BZ), it allows us to use $\{|u_i^{\bm k_0}\rangle\}$ as a basis to approximately solve the Schr\"{o}dinger equation in the framework of the perturbation theory, where $i$ indicates the band index. Once we know the symmetry transformations of these states,  the matrix representation of $\mathcal{H}_{\bm k}$ under this basis can then be determined in form, thus $E_{\bm k}$ and the vector $|u_{\bm k}\rangle$.

Around a high-symmetry point $\bm k_0$, for a symmetry operation $S$ meeting $\gamma \bm k_0 = \bm k_0$, we can rewrite Eq.~\eqref{Hk_final} as
\begin{equation}\label{equ_61_1}
S \mathcal{H}_{\bm k_0+\delta \bm k}S^{-1} = \mathcal{H}_{\bm k_0+\gamma\delta\bm k}.
\end{equation}
First of all, the symmetry operations under consideration have to keep the linear space spanned by $\{|u_i^{\bm k_0}\rangle\}$  invariant. Otherwise, we must either drop some symmetry operations or add more states into the basis. With the representation matrix $D(S)$ for symmetry operations $S$ defined as
\begin{subequations}\label{equ_62_2}
\begin{equation}
  S^u (|u^{\bm k_0}_1\rangle,|u^{\bm k_0}_2\rangle,...)=(|u^{\bm k_0}_1\rangle,|u^{\bm k_0}_2\rangle,...)D(S^u),
\end{equation}
\begin{equation}
  S^a (|u^{\bm k_0}_1\rangle,|u^{\bm k_0}_2\rangle,...)=(|u^{\bm k_0}_1\rangle,|u^{\bm k_0}_2\rangle,...)D(S^a)\mathcal{K},
\end{equation}
\end{subequations}
where we have explicitly marked the symmetry operation $S$ by $u$ for ``unitary'' and $a$ for ``antiunitary'', Eq.~\eqref{equ_61_1} immediately gives
\begin{subequations}\label{equ_all_H}
\begin{equation}
  D(S^u)\mathcal{H}_{\bm k_0}(\gamma^{-1}\delta\bm k)D(S^u)^{-1} = \mathcal{H}_{\bm k_0}(\delta\bm k),
\end{equation}
\begin{equation}
  D(S^a)\mathcal{H}_{\bm k_0}(\gamma^{-1}\delta\bm k)^*D(S^a)^{-1} = \mathcal{H}_{\bm k_0}(\delta\bm k).
\end{equation}
\end{subequations}
where  $\mathcal{H}_{\bm k_0}(\delta\bm k)$ denotes $\mathcal{H}_{\bm k_0+\delta\bm k}$.

\textit{LCAO basis.} In group theory, for a linear space invariant under a group, one can always find a set of basis in which all the vectors can form irreducible representations of that group and are orthogonal with each other. For a general vector, through analyzing its components, one can easily find all other vectors related by the group operations and all the irreducible representations. Since the first-principles calculations can give the atomic-orbital projection of the wavefuction, it would be useful to set up a database containing the representations of all low-level linear combinations of atomic orbitals (LCAOs).

First of all, we ignore the spin. The periodic part of linear combination of atomic orbitals has the form
\begin{align}
u^{\bm k}_{m\alpha} (\bm r) &= \frac{1}{\sqrt{N}}\sum_{\bm R_n}e^{-i\bm k\cdot(\bm r-\bm r_\alpha-\bm R_n)}\phi_m(\bm r-\bm r_\alpha - \bm R_n),
\end{align}
where $\alpha$ and $m$ respectively refers to the $\alpha$-th atom in a unit cell and its $m$-th atomic orbital. Here we also defined $\phi'_m(\bm r)=e^{-i\bm k\cdot \bm r}\phi_m(\bm r)/\sqrt{N}$. It should be noted that if we collect the periodic wavefunctions for all the same atoms in a unit cell and all the atomic orbitals with the same quantum number $\ell$ to span a linear space, it must be invariant under symmetry operations.

To simplify the calculation, we analyze the indices $m$ and $\alpha$ separately. To do that, we need two linear spaces: a complex linear space $V_1$ spanned by $\{\phi'_m(\bm r)\}$ and $V_2=\mathbb{R}^{\alpha_\text{max}}$, where $\alpha_\text{max}$ means the number of atoms in our consideration of a unit cell and the vectors in it indicate the spatial distribution of atoms. Then with the correspondence as 
\begin{align}
&\phi'_m(\bm r)\otimes (\overbrace{0,...,0}^{\alpha-1},1,0,...,0)^T \notag\\
\leftrightarrow& \sum_{\bm R_n}\phi'_m(\bm r-\bm r_\alpha-\bm R_n),
\end{align}
we know that the tensor product of these two spaces $V_1\otimes V_2$ is isomorphic to the space $V_0$ spanned by $\{u^{\bm k}_{m\alpha}(\bm r)\}$ for a certain $\bm k$.
Finally, with the straightforward correspondence
\begin{align}
&S\sum_{\bm R_n}\phi'_m(\bm r-\bm r_\alpha-\bm R_n) \notag\\
\leftrightarrow& S^P\phi'_m(\bm r)\otimes S(\overbrace{0,...,0}^{\alpha-1},1,0,...,0)^T,
\end{align}
where $S^P$ is the translation-deleted part of $S$. One can easily find the resulting state of any symmetry operations on $u^{\bm k}_{m\alpha} (\bm r)$. Above all, the direct product between the represenations of $V_1$ and $V_2$ is that of $V_0$.

When taking the spin into account, we could regard $V$ as $ V_1\otimes V_2\otimes V_\text{spin}$ . In this way, the discussion above can be immediately generalized to the situation with spin, where we must consider the effect of $S^P$ on the spin part.

For different systems, the representations formed by $V_2$ differ from each other, but they are mostly direct sum of pure one-dimensional representations. Thus in the database, we only collect matches between representations of the atomic orbital with or without spin and all one-dimensional representations.

\textit{Expansion with Hermitian matrices.} Considering the hermiticity of $\mathcal{H}_{\bm k_0}(\delta\bm k)$, it can be expanded by $n^2$ linearly independent Hermitian matrices with real coefficients $h(\delta\bm k)$, and then we can obtain $n^2$ independent equations. We denote these Hermitian matrices by $\{B_i,i=1,2,...,n^2\}$. They span a linear space on real number field. Then $\mathcal{H}_{\bm k_0}(\delta\bm k)$ can be expressed as
\begin{equation}
  \mathcal{H}_{\bm k_0}(\delta\bm k) = \sum_i h_i(\delta\bm k)B_i,
\end{equation}
Therefore, the Eq.~\eqref{equ_all_H} can be expressed as
\begin{subequations}\label{equ_123}
\begin{equation}
  \sum_j h_j(\gamma^{-1}\delta\bm k) D(S^u)B_jD(S^u)^{-1} = \sum_i h_i(\delta\bm k)B_i,
\end{equation}
\begin{equation}
  \sum_j h_j(\gamma^{-1}\delta\bm k) D(S^a)B^*_jD(S^a)^{-1} = \sum_i h_i(\delta\bm k)B_i,
\end{equation}
\end{subequations}
Now we define two new matrices $\mathcal{M}$ and $\mathcal{N}$ as follows,
\begin{subequations}
  \begin{equation}
    D(S^u)B_{j}D(S^u)^{-1} = \sum_m \mathcal{M}_{mj}B_m,
  \end{equation}
  \begin{equation}
    D(S^a)B^*_{j}D(S^a)^{-1} = \sum_m \mathcal{M}_{mj}B_m,
  \end{equation}
\end{subequations}
\begin{equation}\label{h_n}
  h_j(\gamma^{-1}\delta\bm k) = \sum_n \mathcal{N}_{nj}h_n(\delta\bm k),
\end{equation}
Comparing the coefficients of $B_i$ on both sides of the Eq.~\eqref{equ_123}, it yields $\sum_n (\mathcal{MN}^T)_{in}h_n(\delta\bm k) = h_i(\delta\bm k)$ for $i=1,2,...,n^2$, that is
\begin{equation}\label{MN_re}
  \mathcal{N}^Th(\delta\bm k) = \mathcal{M}^{-1}h(\delta\bm k)
\end{equation}

On one hand, if the basis $\{|u^{\bm k_0}_i\rangle, i =1,2,...,n\}$ is orthonormal, it is obvious that $D$ are unitary matrices ($D^{-1}=D^\dagger$), which implies that $D(S^u)B_{j}D(S^u)^{-1}$ are vectors in Hermitian matrix space on real number field, Thus $\mathcal{M}$ must be real.  On the other hand, if we define the inner product of Hermitian matrices as $\text{tr}(B_iB_j^\dagger)$. It can be proved that
\begin{align}
  &\text{tr}(D(S^u)B_{i}D(S^u)^{-1}(D(S^u)B_{j}D(S^u)^{-1})^\dagger)\notag\\
  =&\text{tr}(B_{i}B_{j}^\dagger),
\end{align}
which means the inner product remains unchanged (the same for $S^a$). This implies that if $\{B_i\}$ is an orthonormal basis, $\mathcal{M}$ must be orthogonal ($\mathcal{M}^{-1}=\mathcal{M}^T$).

Especially, if the combination of inversion and time reversal symmetry $\mathcal{PT}$ is in consideration, then we have
\begin{equation}
  \mathcal{M}(\mathcal{PT})h(\delta\bm k) = h(\delta\bm k),
\end{equation}
which allows us to simplify the calculation. If we assume the solution of $[\mathcal{M}(\mathcal{PT})-I] X = 0$ is $\{b_i, i=1,2,...,m\}$, where $m<n^2$ always holds, then $h(\delta\bm k)=\sum_ih'_i(\delta\bm k)b_i$  and $H_{\bm k_0}(\delta\bm k) = \sum_ih'_i(\delta\bm k) B'_i$, where we denoted $\sum_j(b_i)_jB_j$ by $B'_i$. Now we can recalculate $\mathcal{M}'$ with $\{B'_i, i=1,2,...,m\}$, but this time only the $m\times m$ submatrix on the top left corner of $\mathcal{M}'$ is meaningful and actually we can only get this submatrix with the incomplete set of Hermitian matrices $\{B'_i\}$.

\textit{Expansion of $h(\delta\bm k)$.} Since $h(\delta\bm k)$ is a polynomial of $\delta\bm k$, we could expand $h(\delta\bm k)$ by a monomial basis of $\delta\bm k$. Suppose $\{f_i(\delta\bm k), i=1,2,...,d\}$ is the basis, then $h(\delta\bm k)$ can be expanded as
\begin{equation}
h_i(\delta\bm k) = \sum_j f_j(\delta\bm k)A_{ji}.
\end{equation}
Then with the definition of matrix $\mathcal{F}$ below,
\begin{equation}
f_i(\gamma^{-1}\delta\bm k) = \sum_j \mathcal{F}_{ji}f_j(\delta\bm k)
\end{equation}
Eq.~\eqref{h_n} gives $\mathcal{F}A=A\mathcal{N}$, and then Eq.~\eqref{MN_re} gives $A=\mathcal{F}A\mathcal{M}^T$. If defining $\alpha$ as a vector from stacking up columns of $A^T$, we obtain
\begin{equation}
(\mathcal{F}\otimes \mathcal{M}-I)\alpha=0.
\end{equation}
Assuming that the solution is $\alpha = \sum_l c_l\tilde{\alpha}(l)$, where $\{\tilde{\alpha}(l)\}$ is a basis of the solution space and $c_l$ are real coefficients, then $A = \sum_l c_l\tilde{A}(l)$, and finally
\begin{equation}
  H_{\bm k_0}(\delta\bm k) = \sum_ih_i(\delta\bm k)B_i = \sum_{ijl}c_lf_j(\delta\bm k){\tilde{A}(l)}_{ji}B_i
\end{equation}
Note that $\mathcal{F}$ is a block diagonal matrix, since polynomials with different orders cannot transform to each other. We could separately calculate $\mathcal{F}^{(n)}$ and $A^{(n)}$ for each Taylor series expansion order $f^{(n)}(\delta\bm k)$. Finally, we stack $f^{(n)}(\delta\bm k)$ and $A^{(n)}$ respectively to get $f(\delta\bm k)$ and $A$, that is
\begin{equation}
f(\delta\bm k) = \begin{pmatrix}
  f^{(0)}(\delta\bm k)\\
  f^{(1)}(\delta\bm k)\\
  f^{(2)}(\delta\bm k)\\
  \vdots
  \end{pmatrix}
\end{equation}
and
\begin{align}
&\tilde{A}= \sum_{l=1}^{l^{(0)}} c_{l}\begin{pmatrix}
  \tilde{A}^{(0)}(l)\\
  0\\
  0\\
  \vdots
\end{pmatrix}
+\sum_{l=l^{(0)}+1}^{l^{(0)}+l^{(1)}} c_{l}\begin{pmatrix}
  0\\
  \tilde{A}^{(1)}(l)\\
  0\\
  \vdots
\end{pmatrix}\notag\\
&+\sum_{l=l^{(0)}+l^{(1)}+1}^{l^{(0)}+l^{(1)}+l^{(2)}}c_{l}\begin{pmatrix}
  0\\
  0\\
  \tilde{A}^{(2)}(l)\\
  \vdots
\end{pmatrix}+...
\end{align}



\textit{The case of $\bm k_0$ on the BZ boundary under a symmetry operation $\gamma $ with $\gamma \bm k_0-\bm k_0\neq0$.} When the $\bm k_0$ is on the BZ boundary, the following equation should be taken into account,
\begin{equation}\label{Hk_plusG}
\mathcal{H}_{\bm k+\bm G} = e^{-i\bm G\cdot \bm r}\mathcal{H}_{\bm k}e^{i\bm G\cdot \bm r},
\end{equation}
where $\bm G$ is an arbitrary reciprocal lattice vector.
With this equation, the symmetry operation meeting $\gamma \bm k_0-\bm k_0=\bm G\neq0$, can also limit the form of the Hamiltonian around $\bm k_0$, since
\begin{align}
&S\mathcal{H}_{\bm k_0+\delta \bm k}S^{-1} = \mathcal{H}_{\gamma\bm k_0+\gamma \bm \delta k} =e^{-i\bm G\cdot \bm r}\mathcal{H}_{\bm k_0+\gamma \delta \bm k}e^{i\bm G\cdot \bm r}.
\end{align}
We can define $\tilde{S}$ as $\tilde{S} = e^{i\bm (\gamma\bm k_0 - \bm k_0)\cdot \bm r}S$, which leads to an equation with the same form as Eq.~\eqref{equ_61_1},
\begin{equation}
\tilde{S} \mathcal{H}_{\bm k_0+\delta\bm k}\tilde{S}^{-1} = \mathcal{H}_{\bm k_0+\gamma\delta \bm k}.
\end{equation}
Although the set of symmetry operations $\{\tilde{S}\}$ actually describes the full symmetry of $\mathcal{H}_{\bm k_0}$, it is not always a group. The product of two arbitrary elements $\tilde{S}_1$ and $\tilde{S}_2$ is
\begin{equation}
\tilde{S}_1\tilde{S}_2 = e^{i\gamma_1(\bm k_0-\gamma_2\bm k_0)\cdot \bm \tau_1}e^{i(\gamma_1\gamma_2\bm k_0-\bm k_0)\cdot \bm r}S_1S_2,
\end{equation}
which is still in $\{\tilde{S}\}$ and corresponding to $S_1S_2$ only if $e^{i\gamma_1(\bm k_0-\gamma_2\bm k_0)\cdot \bm \tau_1}=1$. 
Due to the existence of the symmetry $\gamma$ with $\gamma \bm k_0\neq\bm k_0$, only if all symmetries in consideration are symmorphic ($\tau=0$), $\{\tilde{S}\}$ becomes a group. Otherwise, we either drop the symmetry operation $\gamma$ with $\gamma \bm k_0\neq\bm k_0$ or the nonsymmorphic symmetry operations in our code.

\begin{figure}
\includegraphics[width=8.5cm]{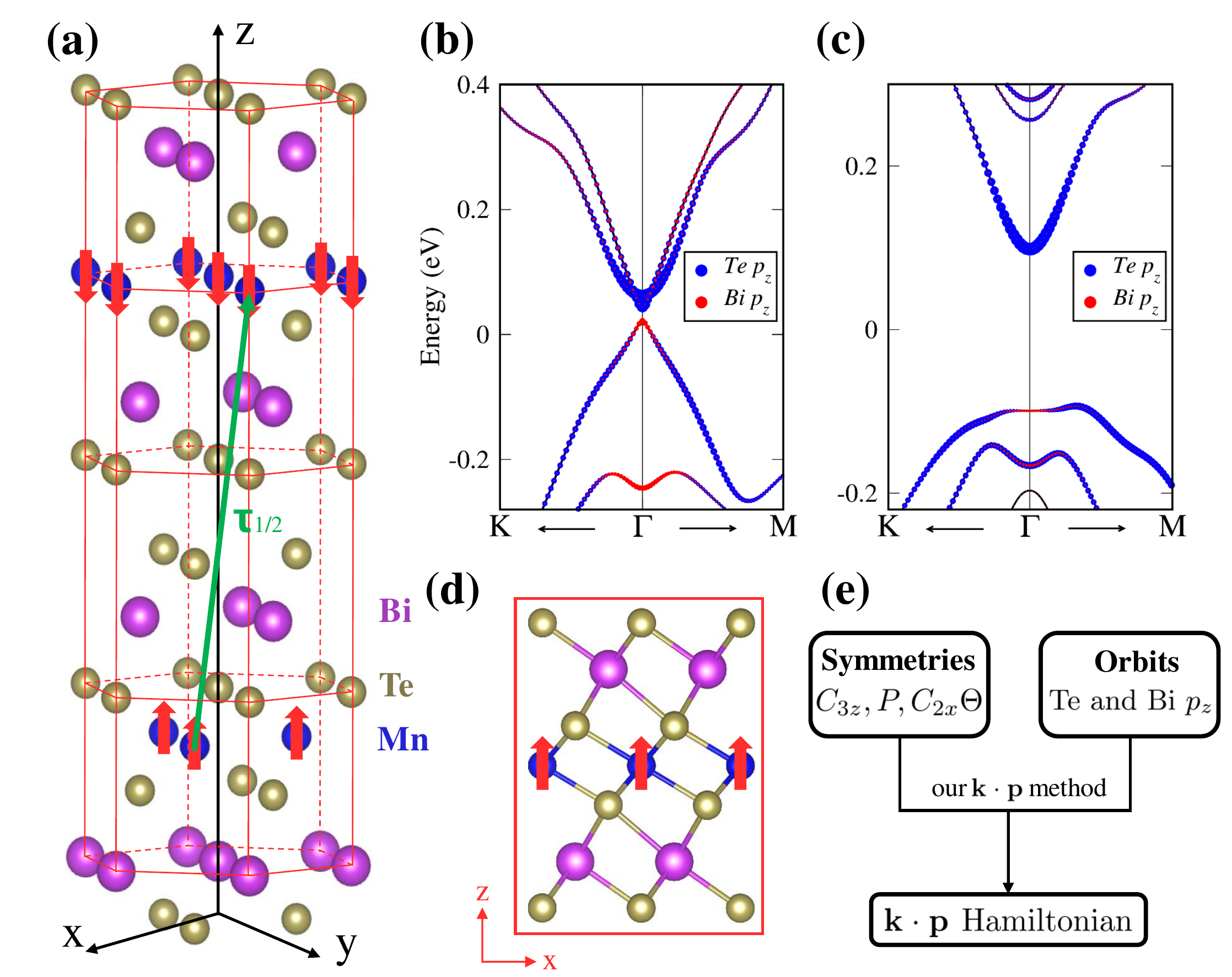}
\caption{\label{fig:fig2}\textbf{Crystal structure and band structure of magnetic topological insulator MnBi$_2$Te$_4$.} (a) The unit cell of AFM MnBi$_2$Te$_4$ consists of two SLs. The red arrows represent the spin moment of Mn atom. The green arrow denotes for the half translation operator $\tau_{1/2}$. (b)(c)The fat band structures of FM and AFM states . The blue and red dots denote the characters of Te and Bi $p_z$-orbitals respectively. (d) The unit cell of FM MnBi$_2$Te$_4$ has one SL. (e) The process of construct FM MnBi$_2$Te$_4$ $\bm{k\cdot p}$ Hamiltonian. }
\end{figure}

\textit{Application to magnetic topological material MnBi$_2$Te$_4$.}  MnBi$_2$Te$_4$ is a versatile magnetic topological materials to realize to quantum anomalous Hall (QAH) state, antiferromagnetic (AFM) topological insulator, magnetic axion insulator, tunable dynamical axion field and ferromagnetic (FM) Weyl semimetal\cite{zhang2019,li2019,gong2019,otrokov2019nature,otrokov2019,deng2020,liu2020robust,chen2019intrinsic,klimovskikh2020npj}. Here, with the above method, we take MnBi$_2$Te$_4$ as an example to demonstrate how to construct the $\bm{k\cdot p}$ Hamiltonian.

As shown in Fig.~\ref{fig:fig2}(a), MnBi$_2$Te$_4$ has a layered crystal structure with a triangle lattice. The trigonal axis (threefold rotation symmetry $C_{3z}$) is defined as the $z$ axis, a binary axis (twofold rotation symmetry $C_{2x}$) is defined as the $x$ axis and a bisectrix axis (in the reflection plane) is defined as the $y$ axis for the coordinate system. The material consists of septuple layers (SL) (e.g., Te-Bi-Te-Mn-Te-Bi-Te) arranged along the $z$ direction. The structural of nonmagnetic MnBi$_2$Te$_4$ is described by the space group $D^5_{3d}$ ($R\bar{3}m$) whose generators are $C_{3z}$, $P$, $C_{2x}$ and pure translations which can be ignored since their representations formed by LCAOs are identities. The first-principles calculations of FM and AFM states(Fig.~\ref{fig:fig2}(b)(c)) show that four bands at the $\Gamma$ point near the Fermi-level all contains $p_z$ orbitals of Bi and Te atom, the basis for both FM and AFM states can be expressed as $\{|\text{Bi}, p_z,\uparrow\rangle^+, i|\text{Te}, p_z,\uparrow\rangle^-, |\text{Bi}, p_z,\downarrow\rangle^+, -i|\text{Te}, p_z,\downarrow\rangle^-\}$, where the superscripts `$+/-$' indicate parities.

For the FM state, the symmetries are three-fold rotation $C_{3z}$, inversion symmetry $P$, and the combination of two-fold rotation and time reversal symmetry $C_{2x}\Theta$, the representation matrices can be obtained by our code:
\begin{equation}
\begin{cases}
D(C_{3z})=\text{exp}(-i\frac{\pi}3\sigma_z\otimes 1_{2\times2})\\
D(P)=1_{2\times2}\otimes\tau_z\\
D(C_{2x}\Theta) = \text{exp}(i\frac\pi2\sigma_z\otimes 1_{2\times2})\mathcal{K}
\end{cases}
\end{equation} where $\mathcal{K}$ is the complex conjugation operator.

By using our $\bm{k\cdot p}$ method, we first get 0-order $\bm k\cdot\bm p$ Hamiltonian around $\Gamma$ point:
\begin{align}
\mathcal{H}(\bm k) = c_1\sigma_0\otimes\sigma_0+c_4\sigma_0\otimes\sigma_z+c_{13}\sigma_z\otimes\sigma_0+c_{16}\sigma_z\otimes\sigma_z\notag\\
\end{align}
where $c_{i}(i=1,4,13,16)$ are coefficients.

Further, we can also get the $\bm k\cdot\bm p$ Hamiltonian up to 2 order as follows,
\begin{equation}
\mathcal{H}_\text{FM}(\bm k)= \begin{pmatrix}
M_1(\bm k) & A_1k_z & 0 & A_2 k_-\\
A_1k_z & M_2(\bm k) & A_4k_- & 0\\
0 & A_4k_+ & M_3(\bm k) & A_3k_z\\
A_2 k_+ & 0 & A_3k_z & M_4(\bm k)
\end{pmatrix},
\end{equation}
where $k_\pm = k_x\pm ik_y$ and $M_i(\bm k) = M_0^i+B_1^ik_z^2+B_2^i(k_x^2+k_y^2)$. Note that the off-diagonal two-order terms are omitted, since they contribute higher order of $\bm k$ to the energy.

To show the change of the Hamiltonian induced by the FM order, we now assume both $C_{2x}$ and $\Theta$ are preserved. In addition to $D(C_{3z}),D(P)$, the representation matrices also include $D(C_{2x}) = \text{exp}[-i(\pi/2)\sigma_x\otimes 1_{2\times2}]$, $D(\Theta)=i\sigma_y\otimes 1_{2\times2}\mathcal{K}$. Then we can get the $\bm k\cdot\bm p$ Hamiltonian of nonmagnetic state
\begin{equation}
\mathcal{H}_\text{NM}(\bm k) = \epsilon_0(\bm k)+\begin{pmatrix}
M_\gamma(\bm k) & A_1k_z &0 & A_2k_-\\
A_1k_z & -M_\gamma(\bm k) & A_2k_- & 0\\
0 & A_2k_+ & M_\gamma(\bm k) & -A_1k_z\\
A_2 k_+ & 0& -A_1k_z & -M_\gamma(\bm k)
\end{pmatrix},
\end{equation}
where $\epsilon_0(\bm k) = C + D_1k_z^2+ D_2(k_x^2+k_y^2)$ and $M_\gamma(\bm k) = M_0^\gamma+B_1^\gamma k_z^2+B_2^\gamma(k_x^2+k_y^2)$. Comparing these two Hamiltonian, we find the perturbative term induced by the FM magnetic structure
\begin{equation}
\delta \mathcal{H}_\text{FM}(\bm k) = \begin{pmatrix}
M_+(\bm k) & A_3k_z &0 &A_4k_-\\
A_3k_z & M_-(\bm k) & A_4k_-&0\\
0 & -A_4 k_+ & -M_+(\bm k) & A_3k_z\\
A_4 k_+ & 0 & A_3k_z & -M_-(\bm k)
\end{pmatrix}
\end{equation}
where $M_\pm(\bm k) = M_\alpha(\bm k)\pm M_\beta(\bm k)$, $M_j(\bm k) = M_0^j+B_1^jk_z^2+B_2^j(k_x^2+k_y^2)$ with $j=\alpha, \beta$.

For the A-type AFM state, the time reversal symmetry $\Theta$ is broken, but the combination operation $\Theta\tau_{1/2}$ of $\Theta$ and a translation $\tau_{1/2}$ (shown in Fig.~\ref{fig:fig2}(a)) is preserved. Compared with nonmagnetic state, the representation of the combination symmetry $\Theta\tau_{1/2}$ in AFM state is the same as that of the time reversal symmetry $\Theta$. Thus their Hamiltonian are the same in form.

\begin{acknowledgments}
This work is supported by the Fundamental Research Funds for the Central Universities (Grant No. 020414380149), Natural Science Foundation of Jiangsu Province (No. BK20200007), the Natural Science Foundation of China (Grants No. 12074181, No. 11674165 and NO. 11834006) and the Fok Ying-Tong Education Foundation of China (Grant No. 161006).

G. Zhan  and M. Shi contributed equally to this work.
\end{acknowledgments}

\appendix
\section{Point Operations on Orbits}
In our method, an important step is getting the representation matrices of a symmetry operation formed by LCAOs. There are some representation matrix under three orbits $s,p,d$.

(1) The $s$ orbital is invariant under any point operations. We denote this representation matrix as $D_s$: $D_s=1$.

(2) The $p$ orbitals $|p_0\rangle$, $|p_{\pm1}\rangle$ can be combined into several states whose transformation under point operations are clearer as follow:
\begin{equation}
\begin{cases}
|p_x\rangle=\frac{1}{\sqrt{2}}(|p_1\rangle+|p_{-1}\rangle)\\
|p_y\rangle=\frac{1}{i\sqrt{2}}(|p_1\rangle-|p_{-1}\rangle)\\
|p_z\rangle=|p_0\rangle
\end{cases}
\end{equation}
It is well-known that $\langle\bm r|p_x\rangle=xF_p(\bm r),\langle \bm r|p_y\rangle=yF_p(\bm r),\langle\bm r|p_z\rangle=zF_p(\bm r)$, which results in
\begin{equation}
S(|p_x\rangle,|p_y\rangle,|p_z\rangle) = (|p_x\rangle,|p_y\rangle,|p_z\rangle)g.
\end{equation}
this $3\times3$ matrix $g$ can be easily calculated by $R_x(\alpha)R_y(\beta)R_z(\gamma)$, and we denote this representation matrix as $D_p$: $D_p=g$.

(3) The $d$ orbitals $|d_0\rangle,|d_{\pm1}\rangle,|d_{\pm2}\rangle$ can also form a new set of states as follows:
\begin{equation}
\begin{cases}
|d_{xz}\rangle=\frac{1}{\sqrt{2}}(|d_1\rangle+|d_{-1}\rangle)\\
|d_{yz}\rangle=\frac{1}{i\sqrt{2}}(|d_1\rangle-|d_{-1}\rangle)\\
|d_{xy}\rangle=\frac{1}{i\sqrt{2}}(|d_2\rangle-|d_{-2}\rangle)\\
|d_{x^2-y^2}\rangle=\frac{1}{\sqrt{2}}(|d_2\rangle+|d_{-2}\rangle)\\
|d_{z^2}\rangle=|d_0\rangle
\end{cases}
\end{equation}
whose wavefunctions are of the form $\langle \bm r|d_{xz}\rangle=2xzF_d(\bm r)$,$\langle\bm r|d_{yz}\rangle=2yzF_d(\bm r)$,$\langle \bm r|d_{xy}\rangle=2xyF_d(\bm r)$,$\langle\bm r|d_{x^2-y^2}\rangle=(x^2-y^2)F_d(\bm r)$,$\langle \bm r|d_{z^2}\rangle=(2z^2-x^2-y^2)/\sqrt{3}F_d(\bm r)$. The representation matrix formed by these five orbitals as $D_{d}$.

When considering the spin, the representation matrices must take account into the spin part: $D = D_{s,p,d}\otimes D_{spin}$. Furthermore, if there exists spin-orbit interaction, the magnetic quantum numbers $m_l$ and $m_s$ are no longer good ones. The states with good quantum number $j$ and $m$ can be obtained by combining $|m_l,m_s\rangle$ with $CG$ coefficients. For example,
\begin{equation}
|\frac{3}{2},\frac{3}{2}\rangle = |p_1,+\rangle = \frac{1}{\sqrt{2}}(|p_x,+\rangle+i|p_y,+\rangle).
\end{equation}
The representation matrix formed by a new basis is just similar to the old: $D_{new}=C^{-1}D_{old}C$, where $C$ is the transition matrix.
\section{\label{appendix:b}MnBi$_2$Te$_4$ $\bm{k}\cdot \bm{p}$ Hamiltonian}
There are more detailed derivation to construct MnBi$_2$Te$_4$ $\bm{k}\cdot \bm{p}$ Hamiltonian in our method.

Firstly, for the FM state, both $C_{2x}$ and $\Theta$ are broken, but their combination $C_{2x}\Theta$ preserves. We find that the top and bottom bands contain the $|p_z, \uparrow\rangle$ and $|p_z, \downarrow\rangle$ orbitals of Bi atoms. Representations in $V_1$ and $V_2$ space are respectively
\begin{equation}
\begin{cases}
C_{3z}=\text{exp}(-i\frac{\pi}3\sigma_z)\\
P=-1_{2\times2}\\
C_{2x}\Theta = \text{exp}(-i\frac{\pi}2\sigma_z)
\end{cases},\hspace{1em}
\begin{cases}
C_{3z}=1_{2\times2}\\
P=-1_{2\times2}\\
C_{2x}\Theta = -1_{2\times2}
\end{cases}
\end{equation}
Thus the representation in $V_1\otimes V_2$ space is
\begin{equation}
\begin{cases}
C_{3z}=\text{exp}(-i\frac{\pi}3\sigma_z)\\
P=1_{2\times2}\\
C_{2x}\Theta = \text{exp}(i\frac{\pi}2\sigma_z)
\end{cases}
\end{equation}
The two bands around the Fermi-energy contains $|p_z, \uparrow\rangle$ and $|p_z, \downarrow\rangle$ orbitals of Te atoms, and representations in $V_1$ and $V_2$ space are respectively
\begin{equation}
\begin{cases}
C_{3z}=\text{exp}(-i\frac{\pi}3\sigma_z)\\
P=-1_{2\times2}\\
C_{2x}\Theta = \text{exp}(-i\frac{\pi}2\sigma_z)
\end{cases},\hspace{1em}
\begin{cases}
C_{3z}=1_{2\times2}\\
P=1_{2\times2}\\
C_{2x}\Theta = 1_{2\times2}
\end{cases}
\end{equation}
Thus the representation in $V_1\otimes V_2$ space is
\begin{equation}
\begin{cases}
C_{3z}=\text{exp}(-i\frac{\pi}3\sigma_z)\\
P=-1_{2\times2}\\
C_{2x}\Theta = \text{exp}(-i\frac{\pi}2\sigma_z)
\end{cases}
\end{equation}
Finally, the representation formed by the four states $\{|\text{Bi}, p_z,\uparrow\rangle^+, i|\text{Te}, p_z,\uparrow\rangle^-, |\text{Bi}, p_z,\downarrow\rangle^+, -i|\text{Te}, p_z,\downarrow\rangle^-\}$ (where the superscripts indicate parities) can be obtained by summing up the two representation and making a similarity transformation:
\begin{equation}
\begin{cases}
D(C_{3z})=\text{exp}(-i\frac{\pi}3\sigma_z\otimes 1_{2\times2})\\
D(P)=1_{2\times2}\otimes\tau_z\\
D(C_{2x}\Theta) = \text{exp}(i\frac\pi2\sigma_z\otimes 1_{2\times2})\mathcal{K}
\end{cases}
\end{equation} where $\mathcal{K}$ is the complex conjugation operator. To construct a $\bm k \cdot \bm p$ Hamiltonian, we firstly need a basis of orthonormal Hermitian matrices. $\Gamma$ matrices are well-known $4\times4$ Hermitian matrices which are pairwise orthogonal and can be adopted directly. This base can be expressed as $\{B_i|i=1,2,...,16\}=\{\sigma_i\otimes\sigma_j | i,j=0,x,y,z\}$, and the $M$ matrices can be obtained in this base as follows,
\begin{equation}
\begin{cases}
\mathcal{M}(C_{3z}) = (1\oplus(-\frac12-i\frac{\sqrt{3}}{2}\sigma_y)\oplus1)\otimes1_{4\times4}\\
\mathcal{M}(P) = (\sigma_z\oplus1_{2\times2})\otimes(1_{2\times2}\oplus -\sigma_z)\\
\mathcal{M}(C_{2x}\Theta) = (\sigma_z\oplus1_{2\times2})\otimes(1_{2\times2}\oplus -\sigma_z)
\end{cases}
\end{equation}
For the 0-order polynomial of $\bm k$, that is ${1}$, we have $\mathcal{F}=1$, then $(\mathcal{F}\otimes \mathcal{M}-I)\alpha=0$ for each operation can be solved, the results are
\begin{equation}
\begin{cases}
\alpha({C_3z}) = \sum_{l=1,2,3,4,13,14,15,16}c_lv(l)\\
\alpha({P}) = \sum_{l=1,4,5,8,9,12,13,16}c_lv(l)\\
\alpha({C_{2x}\Theta}) = \sum_{l=1,2,4,9,11,13,14,16}c_lv(l)
\end{cases}
\end{equation}
where $c_l$ are any real number, and
\begin{equation}
v(l)=(\underbrace{0,...,0}_{l-1},1,0,...,0)^T.
\end{equation}
By solving the intersection of these three solutions we find
\begin{equation}
\tilde{A} = \sum_{l=1,4,13,16}c_lv(l)^T,
\end{equation}
and finally we get the 0-order $\bm k\cdot\bm p$ Hamiltonian around $\Gamma$ point
\begin{align}
\mathcal{H}(\bm k) =& \sum_{i}\sum_{l=1,4,13,16}c_lv(l)_{i1}B_i\notag\\
=& c_1\sigma_0\otimes\sigma_0+c_4\sigma_0\otimes\sigma_z+c_{13}\sigma_z\otimes\sigma_0+c_{16}\sigma_z\otimes\sigma_z\notag\\
\end{align}
Similarly, we can get the $\bm k\cdot\bm p$ Hamiltonian up to 2 order as follows,
\begin{equation}
\mathcal{H}_\text{FM}(\bm k)= \begin{pmatrix}
M_1(\bm k) & A_1k_z & 0 & A_2 k_-\\
A_1k_z & M_2(\bm k) & A_4k_- & 0\\
0 & A_4k_+ & M_3(\bm k) & A_3k_z\\
A_2 k_+ & 0 & A_3k_z & M_4(\bm k)
\end{pmatrix},
\end{equation}
where $k_\pm = k_x\pm ik_y$ and $M_i(\bm k) = M_0^i+B_1^ik_z^2+B_2^i(k_x^2+k_y^2)$. Note that the off-diagonal two-order terms are omitted since they contribute higher order of $\bm k$ to the energy. To show the change of the Hamiltonian induced by the broken of symmetry, we now assume both $C_{2x}$ and $\Theta$ are preserved. Their representations for top and bottom bands in $V_1$ and $V_2$ space are respectively
\begin{equation}
\begin{cases}
C_{2x} = \text{exp}(i\frac\pi2\sigma_x)\\
\Theta = i\sigma_y
\end{cases},\hspace{1em}
\begin{cases}
C_{2x} = -1_{2\times2}\\
\Theta = 1_{2\times2}
\end{cases}
\end{equation}
and for the conduction band and valence band are
\begin{equation}
\begin{cases}
C_{2x} = \text{exp}(i\frac\pi2\sigma_x)\\
\Theta = i\sigma_y
\end{cases},\hspace{1em}
\begin{cases}
C_{2x} = 1_{2\times2}\\
\Theta = 1_{2\times2}
\end{cases},
\end{equation}
where the matrices of $C_{3z}$ and $P$ are omitted since they keep unchanged.
Thus representation matrices are $C_{2x} = \text{exp}(-i(\pi/2)\sigma_x\otimes 1_{2\times2}]$, $\Theta=i\sigma_y\otimes 1_{2\times2}K$. We finally get the $\bm k\cdot\bm p$ Hamiltonian of nonmagnetic state
\begin{equation}
\mathcal{H}_\text{NM}(\bm k) = \epsilon_0(\bm k)+\begin{pmatrix}
M_\gamma(\bm k) & A_1k_z &0 & A_2k_-\\
A_1k_z & -M_\gamma(\bm k) & A_2k_- & 0\\
0 & A_2k_+ & M_\gamma(\bm k) & -A_1k_z\\
A_2 k_+ & 0& -A_1k_z & -M_\gamma(\bm k)
\end{pmatrix},
\end{equation}
where $\epsilon_0(\bm k) = C + D_1k_z^2+ D_2(k_x^2+k_y^2)$ and $M_\gamma(\bm k) = M_0^\gamma+B_1^\gamma k_z^2+B_2^\gamma(k_x^2+k_y^2)$. Comparing these two Hamiltonian, we find the perturbative term induced by the FM magnetic structure
\begin{equation}
\delta \mathcal{H}_\text{FM}(\bm k) = \begin{pmatrix}
M_+(\bm k) & A_3k_z &0 &A_4k_-\\
A_3k_z & M_-(\bm k) & A_4k_-&0\\
0 & -A_4 k_+ & -M_+(\bm k) & A_3k_z\\
A_4 k_+ & 0 & A_3k_z & -M_-(\bm k)
\end{pmatrix}
\end{equation}
where $M_\pm(\bm k) = M_\alpha(\bm k)\pm M_\beta(\bm k)$, $M_j(\bm k) = M_0^j+B_1^jk_z^2+B_2^j(k_x^2+k_y^2)$ with $j=\alpha, \beta$.

For the AFM state, $\Theta$ is broken, but the combination of $\Theta$ and a translation $\tau_{1/2}$ (shown in Fig. \ref{fig:fig2}(d)), that $\Theta\tau_{1/2}$ is preserved. The two bands below Fermi energy contain $|p_z, \uparrow\rangle$ and $|p_z, \downarrow\rangle$ orbitals of Bi atoms and their representation in $V_2$ space is
\begin{equation}
\begin{cases}
C_{3z}=1_{2\times2}\\
P = -1_{2\times2}\\
C_{2x} = -1_{2\times2}\\
\Theta\tau_{1/2} = 1_{2\times2}
\end{cases}
\end{equation}
whereas the two bands above the Fermi energy contains $|p_z, \uparrow\rangle$ and $|p_z, \downarrow\rangle$ orbitals of Te atoms, and their representation in $V_2$ space is
\begin{equation}
\begin{cases}
C_{3z}=1_{2\times2}\\
P = 1_{2\times2}\\
C_{2x} = 1_{2\times2}\\
\Theta\tau_{1/2} = 1_{2\times2}
\end{cases}
\end{equation}
As we can see, they are the same as that of the nonmagnetic state. Since the representations in $V_1$ space are obviously the same too, the Hamiltonian expression must also be the same.
\bibliographystyle{apsrev4-1}
\bibliography{kp}
\end{document}